\documentclass[a4paper,10pt]{article}

\usepackage[utf8x]{inputenc}
\usepackage{geometry}
\usepackage{mathrsfs}
\usepackage{graphics, epsfig, subfigure}
\usepackage{amssymb}
\usepackage{fancyhdr} 
\usepackage[english]{babel}
\usepackage{ae}
\usepackage{subfigure} 
\usepackage{amsmath} 
\usepackage{amsfonts}
\usepackage{graphicx}
\usepackage{textcomp}
\usepackage{color} 
\usepackage{float}
\usepackage{lscape}
\usepackage{gensymb} 
\usepackage{authblk} 
\usepackage{hyperref} 

\renewcommand{\exp}[1]{\text{Exp}\left[#1\right]}
\DeclareMathOperator{\nth}{n_{th}}
\newcommand{\txt}[1]{\text{\tiny{#1}}} 

\graphicspath{{../images/}}
\title{Ab-initio Quantum Enhanced Optical Phase Estimation Using Real-time Feedback Control}
\author[1]{Adriano A. Berni}
\author[1]{Tobias Gehring}
\author[1]{Bo M. Nielsen}
\author[2]{Vitus H\"{a}ndchen}
\author[3]{Matteo G.A. Paris}
\author[1]{Ulrik L. Andersen\footnote{Corresponding Author (ulrik.andersen@fysik.dtu.dk)}}
\affil[1]{\textit{\small Department of Physics, Technical University of Denmark, Fysikvej, 2800 Kgs. Lyngby, Denmark}}
\affil[2]{\textit{\small Max-Planck-Institut f\"{u}r Gravitationsphysik (Albert-Einstein-Institut) and Institut f\"{u}r Gravitationsphysik, Leibniz Universit\"{a}t Hannover, Callinstra\ss e 38, 30167 Hannover, Germany}}
\affil[3]{\textit{\small Department of Physics, Universit\`{a} degli Studi di Milano, I-20133 Milano, Italy}}

\date{\today}

\begin{document}
\maketitle
\setcounter{page}{1}

\textbf{
Optical phase estimation is a vital measurement primitive that is used to perform accurate measurements of various physical quantities like length, velocity and displacements~\cite{Caves1981,Giovannetti2011}. The precision of such measurements can be largely enhanced by the use of entangled or squeezed states of light as demonstrated in a variety of different optical systems~\cite{Nagata2007,Banaszek2009,Kacprowicz2010,Taylor2013,GEO,Yonezawa2012,Hoff2013}. Most of these accounts however deal with the measurement of a very small shift of an already known phase, which is in stark contrast to ab-initio phase estimation where the initial phase is unknown~\cite{Xiang2010,Wiseman1995,Berry2006,Kolodynski2010}. Here we report on the realization of a quantum enhanced and fully deterministic phase estimation protocol based on real-time feedback control. Using robust squeezed states of light combined with a real-time Bayesian estimation feedback algorithm, we demonstrate deterministic phase estimation with a precision beyond the quantum shot noise limit. The demonstrated protocol opens up new opportunities for quantum microscopy, quantum metrology and quantum information processing.}

Parameter estimation is an integral part of any physical experiment. In some cases the parameter under interrogation can be measured sharply and thus the uncertainty associated with the measurement is solely governed by the fluctuations of the parameter itself. In other cases, a sharp, canonical measurement is not realizable even in principle. The optical phase is an example of such a parameter~\cite{Susskind1964,Carruthers1968,Barnett1986}. Due to the immense importance in performing an accurate phase measurement in imaging, metrology and communication, numerous theoretical proposals on designing optimised phase measurements have been put forward. The basic aim is to devise a scheme that achieves the sharpest probability distribution for the phase measurement given a fixed amount of resources. 
                     
Quantum estimation theory provides ultimate bounds on the variance of such probability distribution \cite{Hradil2005,Escher2011} in the form of the Cramer-Rao theorems \cite{Cramer1946,Helstrom1979,Braunstein1994}: given $N$ probe states, the variance of any unbiased estimator $\hat{\phi}$ is bounded from below by the quantities: 
\begin{equation}\label{Math_CRbounds}
\Delta^2\phi\geq\frac{1}{NF(\phi)}\geq\frac{1}{NH}.
\end{equation}
Here, $F(\phi)$ is the Fisher information (FI), which is a measure of the phase information associated with a certain detection strategy, while the {\it quantum} Fisher information (QFI), $H$, is the maximized FI over all possible detection strategies. The ultimate lower bound on the variance $1/NH$ is called the {\it quantum} Cramer-Rao (QCR) bound whereas the variance $1/NF(\phi)$ is known as the Cramer-Rao (CR) bound. 

Employing coherent states of light, the QFI is proportional to the mean number of photons; $H=4\langle n\rangle$ and thus the QCR bound is given by $V=1/4N\langle n\rangle$ - the so-called shot noise limit. This limit is surperior to the standard quantum limit (SQL) which is $V=1/2N\langle n\rangle$ and realized with a heterodyne detector \cite{Dariano1996}. The SQL has been beaten in previous experiments using adaptive measurements of a coherent state~\cite{Armen2002,Wheatley2010}.  

Using non-classical resources, the estimation sensitivity can be greatly enhanced beyond the shot noise limit, and eventually reach the optimal Heisenberg scaling for which $V\propto 1/N\langle n\rangle^2$. One class of quantum states of particular interest is the class of Gaussian states as they are relatively easy to produce and comparatively robust against losses \cite{Weedbrook2012}. It has been shown that by employing pure Gaussian squeezed states, the shot noise limit can be beaten and the QCR bound can be asymptotically approached by means of simple homodyne detection, Bayesian estimation and optical feedback \cite{Monras2006,Olivares2009}. In this Letter we demonstrate such a squeezing enhanced quantum phase algorithm. 

\begin{figure*}
\centering
\includegraphics[width=\textwidth]{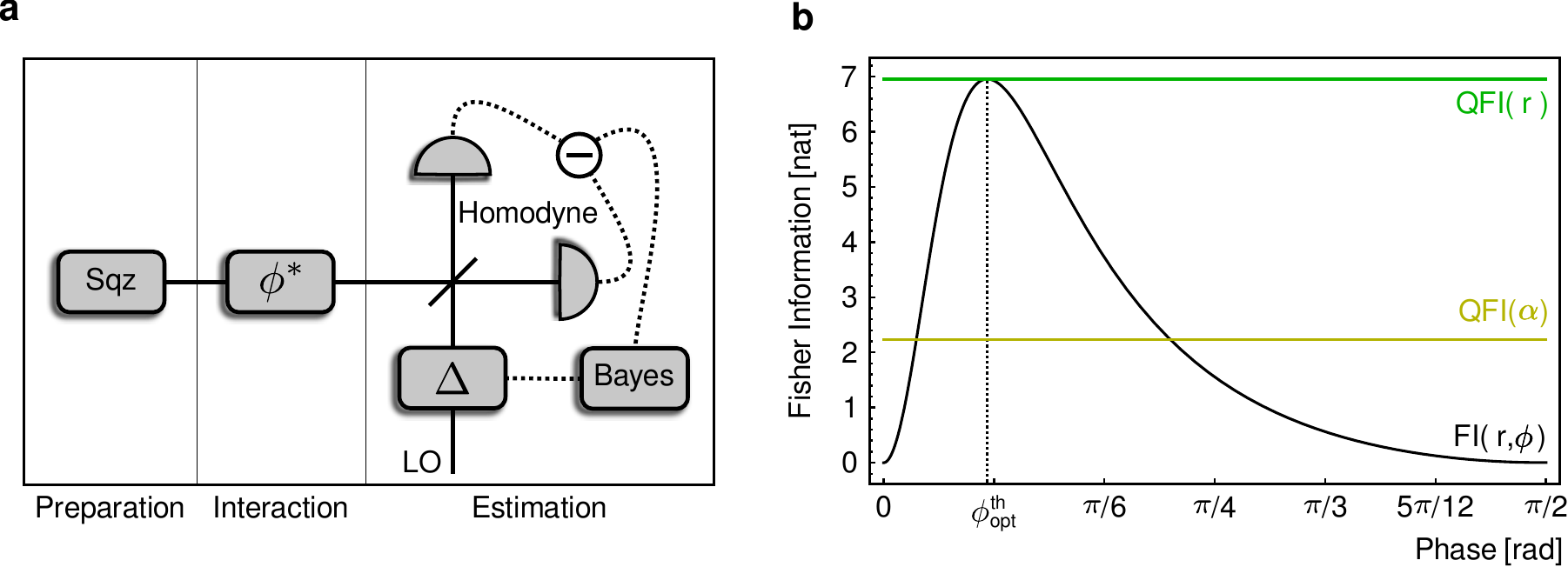}
\caption{\textbf{Principle of squeezing enhanced phase estimation.} \textbf{a)} The input squeezed vacuum state accumulates a phase shift $\phi^*$ which is measured using homodyne detection, real-time Bayesian inference and feedback. The feedback is applied to the local oscillator of the homodyne detector to drive the measuring phase towards the phase which maximizes the Fisher information, $\phi_\txt{opt}^\txt{th}$. \textbf{b)} Fisher information (FI) as a function of the phase for a pure 6dB squeezed vacuum state (black curve). The horizontal lines represent the quantum Fisher information (QFI) for a similarly squeezed state (green) and for a coherent state (yellow). The dashed vertical line indicates the optimal phase.}\label{Fig_Schematic}
\end{figure*}

The phase estimation protocol is schematically shown in Fig.~\ref{Fig_Schematic}a: A squeezed state of light acquires an unknown phase shift within the interval $[0,\pi/2]$ which is subsequently estimated using homodyne detection and Bayesian inference. Using such a measurement, the FI reaches the QFI only for one specific phase shift, $\phi_\text{opt}$, as illustrated in Fig.~\ref{Fig_Schematic}b. This means that by solely using homodyne detection (and Bayesian inference), optimal estimation at the QCR bound is attainable only at one specific phase, and estimation beyond the SQL can be realized only in a limited phase interval. This is sufficient for quantum phase sensing (where a tiny shift of a known phase is measured) but not for ab-initio phase estimation. To circumvent this limitation, the idea is to implement an adaptive feedback scheme such that the system is driven towards the optimal phase. 
More specifically, the strategy is to detect a small fraction of the available samples, use Bayesian inference to obtain a first rough estimate of the phase, employ that information to shift the local oscillator towards the optimal phase point and finally measure the remaining samples to deduce the final estimate via Bayesian inference.  

Bayesian inference provides a framework for deducing the posterior probability distribution (PPD) of the phase conditioned on the M sampled homodyne data $\left\{x\right\}_\txt{M}$:
\begin{equation}\label{Math_PPD}
P(\phi|\left\{x\right\}_\txt{M})\overset{\txt{M}\gg1}{\simeq}\frac{1}{\mathcal{N}}\prod_{i=1}^{\txt{M}}P(x_i,\phi)^{MP(x_i,\phi)},
\end{equation}
where $\mathcal{N}$ is a normalization constant. $P(x_i,\phi)$ are the individual marginal phase distributions conditioned on single homodyne measurement outcomes $x_i$, and is given by
\begin{equation}\label{Math_MPD}
P(x_i,\phi)=\frac{1}{\sqrt{2\pi\sigma_\phi^2}}\exp{-\frac{x_i^2}{2\sigma_\phi^2}},
\end{equation}
where $\sigma_\phi^2=\left(2\nth+1\right)\left[e^{-2r}\cos^2(\phi)+e^{2r}\sin^2(\phi)\right]$ is the variance of the probe state with $r$ the squeezing parameter and $\nth$ the number of thermal photons. Due to the Laplace-Bernstein-Von Mises theorem \cite{LeCam1986}, the posterior distribution of an unknown parameter is independent on the a priori distribution for a large number of homodyne samples, and thus converging to a Gaussian distribution centered on the true value of the parameter, $\phi^\ast$, with a variance $(NF(\phi^\ast))^{-1}$. If $\nth=0$, the use of homodyne detection and Bayesian inference yields operation at the QCR bound, which means that the protocol is optimal.   

Optimality is however conditioned on the state being pure. Adding energy to the state either in terms of incoherent (phase sensitive) noise or in terms of a coherent phase space displacement will indeed increase the phase information (that is, the FI), but the maximized FI for homodyne detection and Bayesian inference is nevertheless only attainable by fuelling the pure squeezing process with all the available energy (see supplementary information). In any realistic implementation, the squeezed state will not be pure but polluted by noise, and thus the actual implementation will not reach the QCR bound but instead a sub-optimal bound that we coin the optimal Cramér-Rao (OCR) bound \cite{Aspachs2009}. 

\begin{figure*}
\centering
\includegraphics[width=\textwidth]{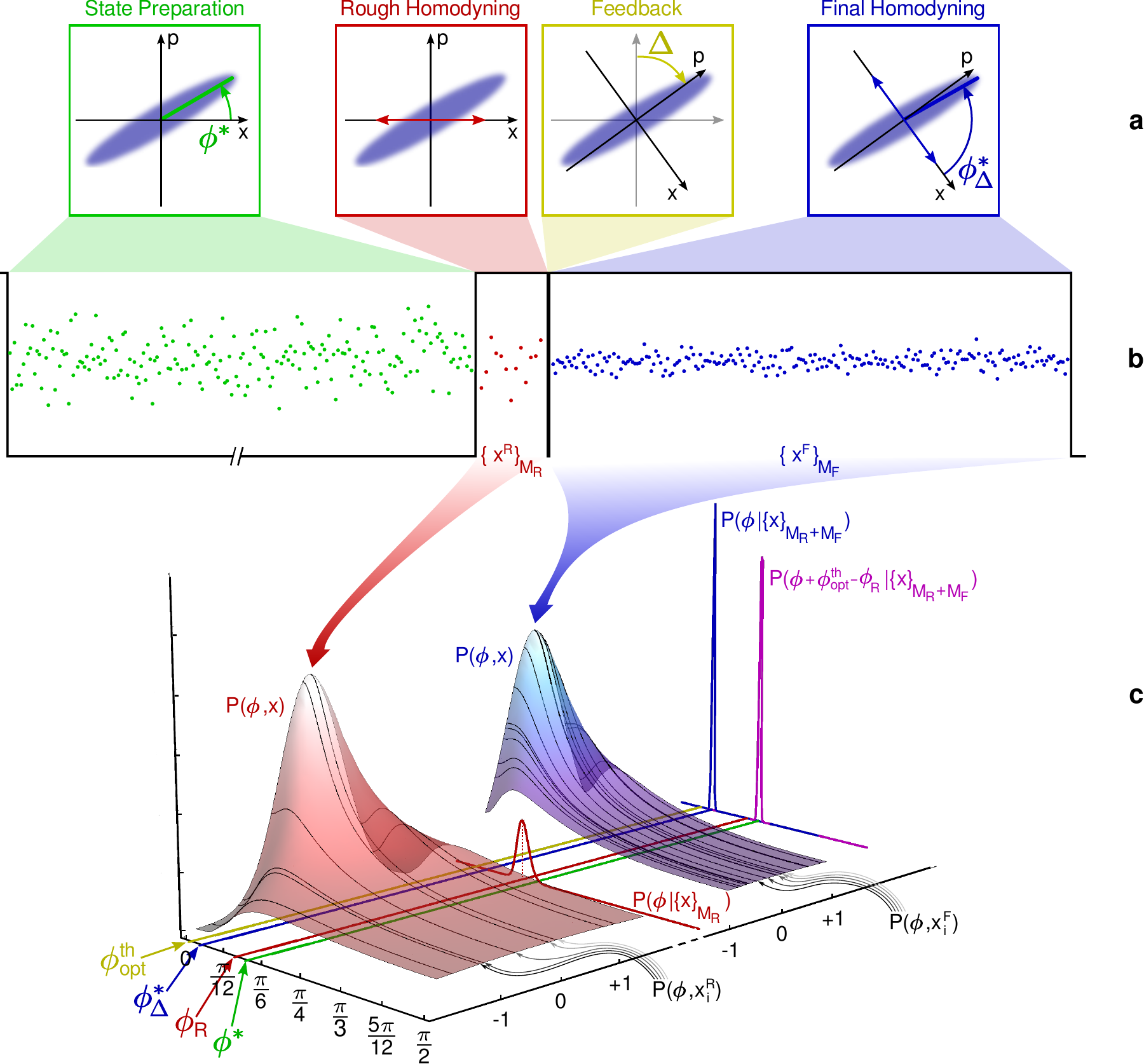}
\caption{\textbf{Working principle of the Bayesian feedback scheme.} \textbf{(a)}: Phase space representation. The protocol starts with state preparation (green), in which a known squeezed thermal state accumulates an unknown phase shift $\phi^\ast$ in the $\left(0,\pi/2\right]$ range. A first detection stage (red) consists of $\text{M}_\txt{R}$ homodyne samples which are used to obtain a rough estimation of the phase shift via Bayesian inference, $\overline{\phi}_\txt{R}$. The rough estimation result is used to compute the matching phase shift, $\Delta$, which is applied to the local oscillator to change the relative phase to the optimal phase $\phi_\txt{opt}^\txt{th}$ (yellow). A second detection stage follows, in which $\text{M}_\txt{F}\gg\text{M}_\txt{R}$ homodyne samples are collected to obtain the final estimation (blue). \textbf{(b)}: Example of homodyne quadrature data for the different steps as a function of time. \textbf{(c)}: Example of an instance of computation of the posterior probability distributions. Each homodyne sample $x_i$ collected during the estimation stages is used to compute a marginal phase distribution $P(\phi,x_i)$. The marginal distributions are multiplied according to Eq.~(\ref{Math_PPD}) to obtain the posterior probability distribution $P(\phi|\left\{x\right\}_\text{M})$ for the rough (left) and the final (right) estimation. The input phase $\phi^\ast$, rough estimation $\overline{\phi}_\txt{R}$, theoretical optimal phase $\phi_\txt{opt}^\txt{th}$ and experimental optimal phase $\phi_\txt{opt}^\txt{exp}$ are shown. We show the resulting PPDs versus phase for a single application of the protocol.
}\label{Fig_Principle}
\end{figure*}

The working principle of the Bayesian scheme combined with adaptive feedback is illustrated in Fig.~\ref{Fig_Principle}. 
We assume that the input state is prepared with an unknown phase shift $\phi^\ast$. The first homodyning stage produces a list of $\text{M}_\txt{R}$ quadrature measurement outcomes each of which is associated with a probability distribution for the phase (see Eq.~(\ref{Math_MPD})). Based on all these distributions, the PPD is computed using the expression in Eq.~(\ref{Math_PPD}) and the rough phase estimate is given by the maximum value, $\overline{\phi}_\txt{R}=\phi^\ast+\delta\phi_\txt{R}$, of the distribution with an estimation error $\delta\phi_\txt{R}$ given by its width. This result is then used to introduce a phase shift, $\Delta=\overline{\phi}_\txt{R}-\phi_\txt{opt}^\txt{th}$, to the local oscillator, thereby shifting the squeezed state phase: $\phi^\ast\longrightarrow\phi^\ast-\Delta=\phi_\txt{opt}^\txt{th}-\delta\phi_\txt{R}=\phi_\txt{opt}^\txt{exp}$. The second Bayesian homodyning stage is then performed, which delivers an estimation of the new phase $\overline{\phi}_\txt{opt}^\txt{exp}=\phi_\txt{opt}^\txt{exp}+\delta\phi_\txt{F}=\phi_\txt{opt}^\txt{th}-\delta\phi_\txt{R}+\delta\phi_\txt{F}$, based on a larger number of samples $\text{M}_\txt{F}\gg\text{M}_\txt{R}$. Due to the larger number of homodyne samples and the enhanced homodyne sensitivity (as a result of the proximity to the optimal phase), we get $\delta\phi_\txt{F}\ll\delta\phi_\txt{R}$. To obtain the final estimate we add $\Delta$ to $\overline{\phi}_\txt{opt}$ and get $\text{Est}\left[\phi^\ast\right]=\overline{\phi}_\txt{opt}^\txt{exp}+\Delta=\phi^\ast+\delta\phi_\txt{F}$, which yields the input phase $\phi^\ast$ with an error $\delta\phi_\txt{F}$.

\begin{figure*}
\centering
\includegraphics[width=\textwidth]{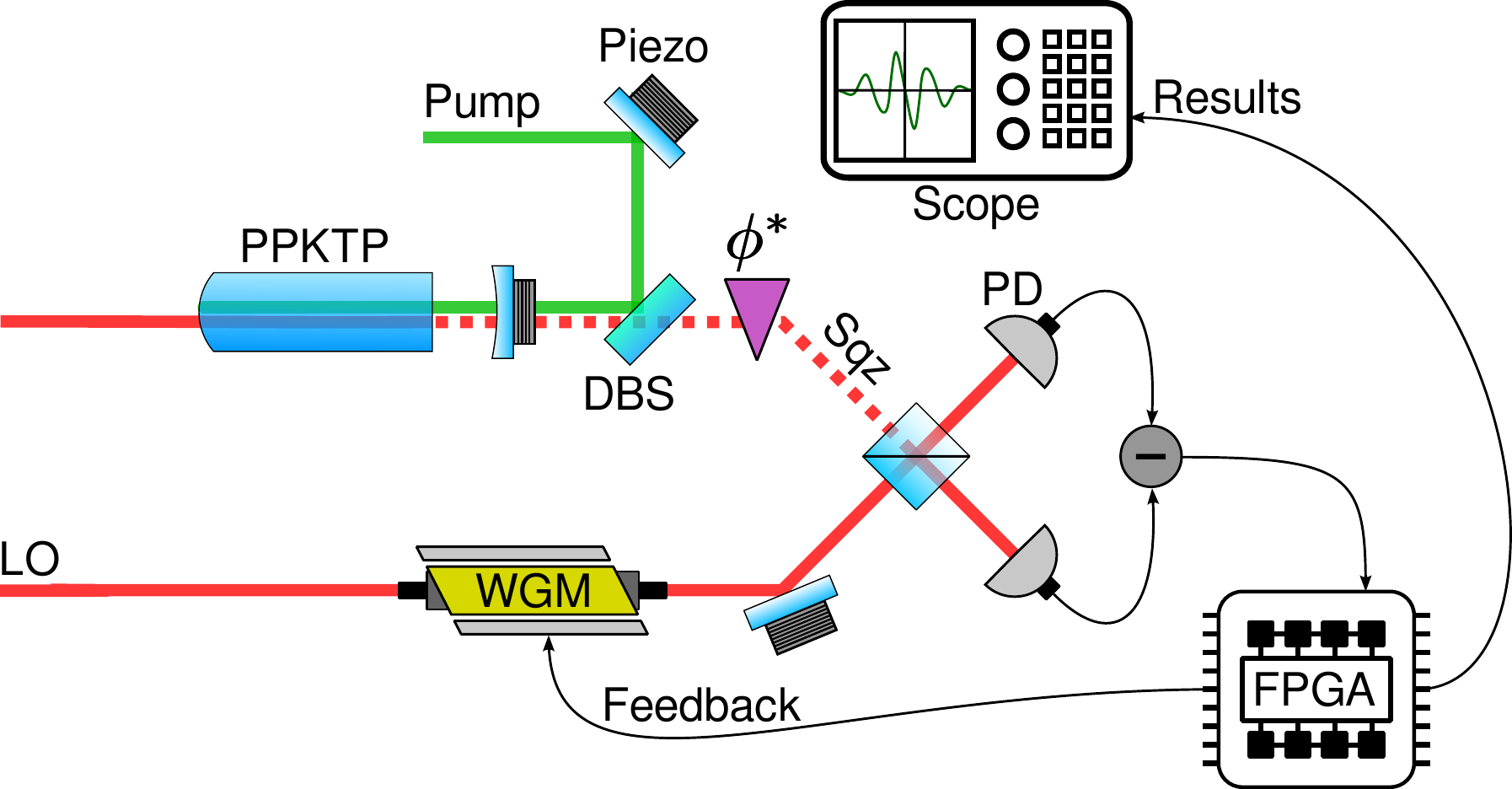}
\caption{\textbf{Simplified experimental layout.} Squeezed vacuum states are generated by degenerate parametric down-conversion using a periodically poled (PP) KTP crystal inside a cavity which is formed by an external mirror and the curved end-facet of the crystal. The process is driven by a pump beam at 775nm producing squeezed light at sidebands around the carrier of 1550nm. The squeezed light is combined with a phase-controlled local oscillator (LO), detected with high-efficiency photodiodes and processed in the FPGA. DBS: dichroic beam splitter. SQZ: squeezed light. PD: photo detector. WGM: waveguide phase modulator. FPGA: field programmable gate array. More information about the setup can be found in the supplementary information.}\label{Fig_Setup}
\end{figure*}

We now turn to the experimental demonstration of the scheme which comprises a source of squeezed light, a homodyne detector and fast feedback electronics. The squeezed light source is based on cavity-enhanced parametric down conversion (see Figure~\ref{Fig_Setup}), and the noise suppression of one quadrature is measured to $-5.69\pm0.07$ dB relative to the vacuum noise limit while the noise of the conjugate quadrature is amplified by $11.83\pm0.09$ dB. These values for the squeezing and anti-squeezing correspond to an energy of $n=3.30\pm0.07$.  

\begin{figure*}
\centering
\includegraphics[width=\textwidth]{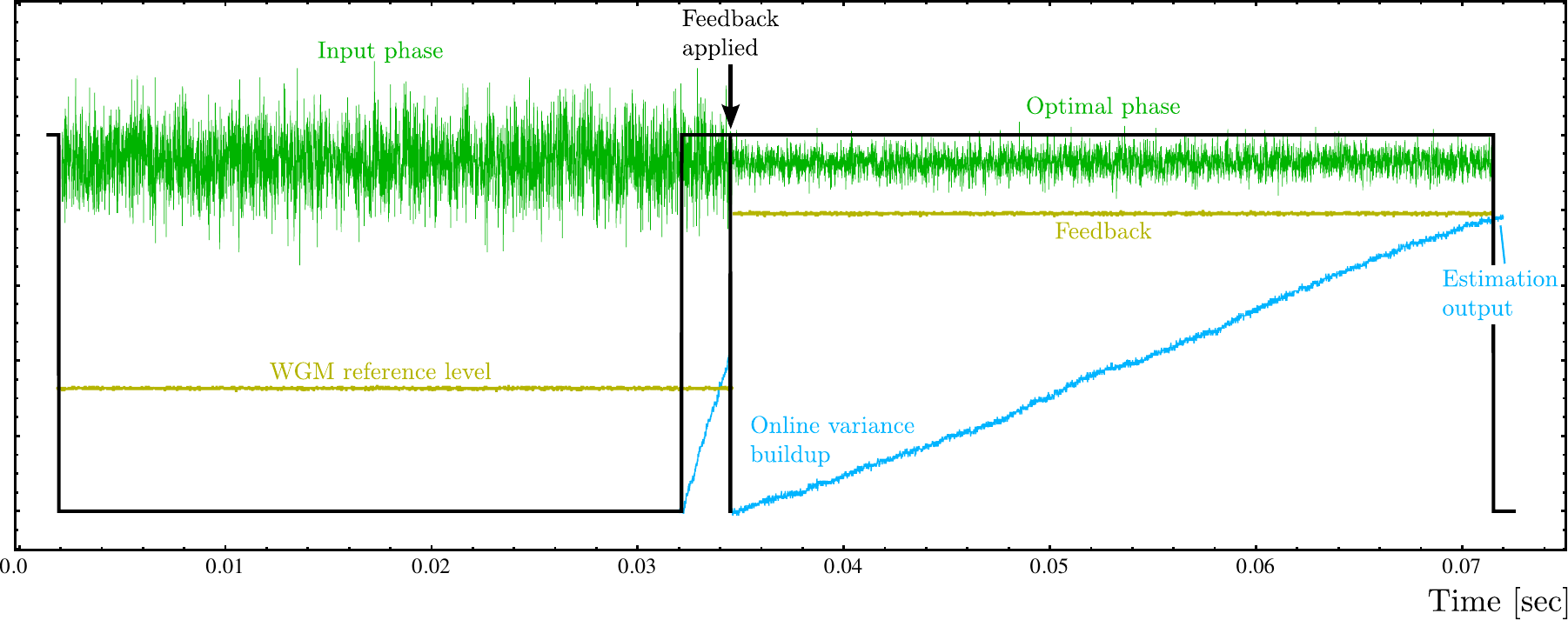}
\caption{\textbf{Illustration of a single measurement sequence.} 
Acquired homodyne quadrature data (green) together with the reference levels (yellow) for the WGM (which controls the relative phase) as a function of time for three different stages: Preparation, rough estimation and final estimation stage. 
During the rough estimation stage the LO is still set on the input phase, thus the homodyne variance is the same as during the state preparation stage. In the final estimation stage, the feedback signal to the WGM is applied (black arrow), which shifts the phase of the state towards the optimal phase, resulting in a change of the spread of the homodyne data. The blue data refer to the on-line accumulation of the (unnormalized) phase variance provided by the FPGA. The final output yields the phase variance.    
}\label{Fig_Signals}
\end{figure*}

For the measurements we use a high-efficiency homodyne detector, including a local (reference) oscillator transversing a fast waveguide phase modulator (WGM). The demodulated homodyne AC signal is fed to a field-programmable gate array (FPGA), which is used to perform the rough estimation based on Bayesian inference as outlined above, and subsequently handles the feedback signal to the waveguide modulator. Due to the FPGA architecture, full implementation of the PPD computation was not convenient, since it would represent a speed bottle neck during the feedback stage. We therefore resorted to an equivalent formalism based on lookup tables and proper calibration (see the supplementary material). 
The estimation result, the homodyne AC signal and the feedback signal are collected by an oscilloscope and a typical output during a single measurement period is shown in Fig.~\ref{Fig_Signals}. The first part of the trace represents state preparation in which the local oscillator is phase locked to the squeezed state. This defines the reference and therefore the input phase (see Fig.~\ref{Fig_Principle}a). Once the estimation stage starts, the phase lock is released to enable the feedback of the FPGA, while after estimation, the lock is reestablished to prepare for the new measurement run. Due to the high stability of the setup, the relative phase between the local oscillator and the squeezed state was not drifting within a single estimation period that lasted between 10 and 70 $\mu$sec.   

\begin{figure*}
\centering
\includegraphics[width=.7\textwidth]{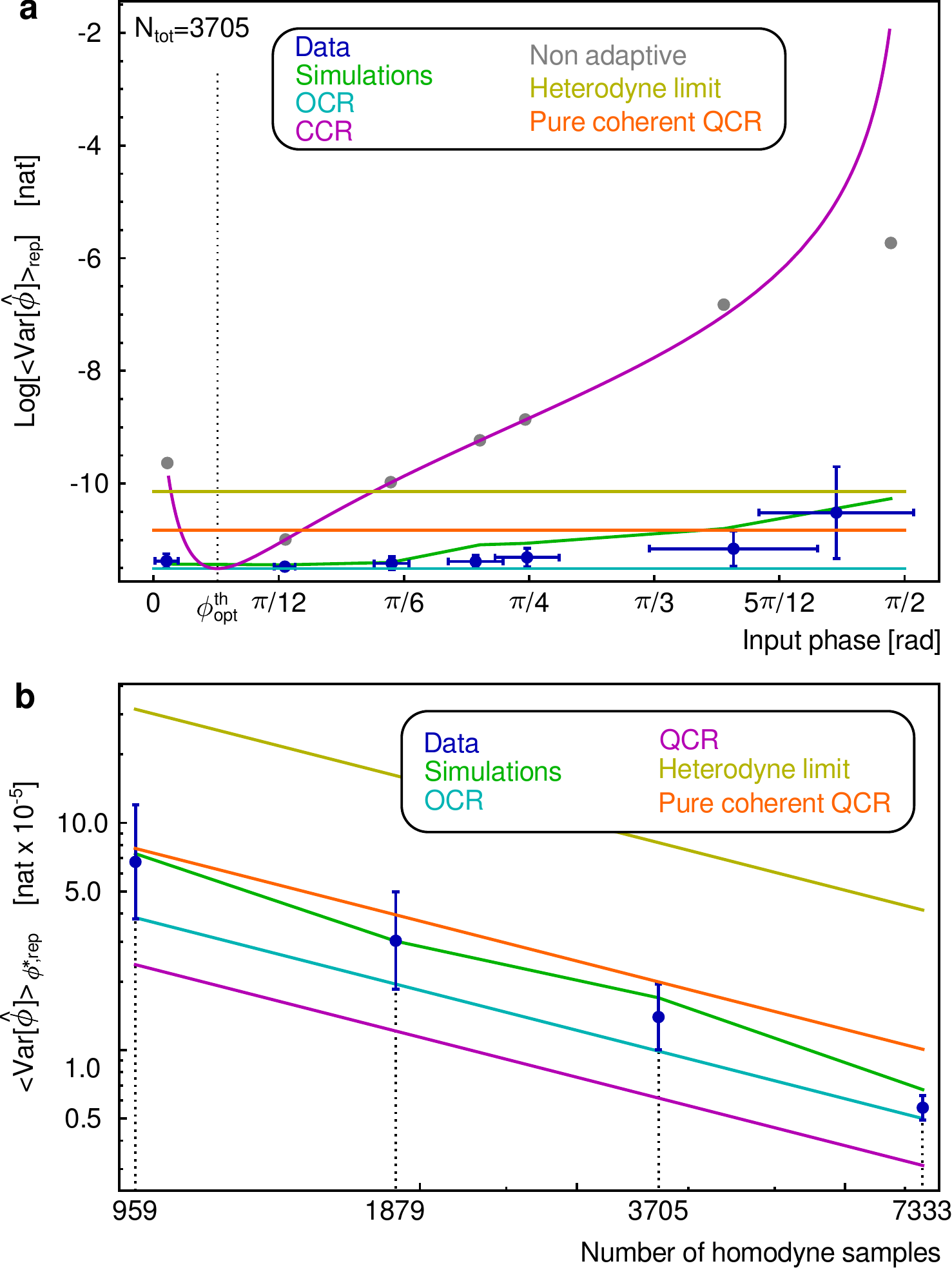}
\caption{\textbf{Estimation variance results.} \textbf{a)} Estimation variance versus input phase for fixed number of homodyne samples. The experimental data (dark blue) are shown for 7 input phases in the $\left(0,\pi/2\right]$ range  with vertical error bars given by the standard deviation over the 80 repetitions and the horizontal error bars given by the standard deviation over the slightly varying input phases over the experimental repetitions. 
\textbf{b)} Estimation variance versus the number of homodyne samples. The experimental results (dark blue) are obtained by averaging the final estimation PPD variance over 80 repetitions of the experiment, and over all tested input phases in each fixed-energy run of the experiment. The error bars are given by the statistical error over the repetitions, averaged over the input phases. The adaptive protocol beats the heterodyne limits and the shot noise limit. 
}\label{Fig_Results}
\end{figure*}

To investigate the performance of the protocol for one specific input phase, we fix the total number of homodyne samples to $\text{N}_\txt{tot}$ and perform 80 repetitions. 
The results of the estimation variance is shown in Fig.~\ref{Fig_Results} (top) for different input phases  together with the theoretical predictions. The experimental results are in good agreement with the expected variance scaling, obtained by simulating the protocol with the same experimental parameters (see the supplementary material). As expected, the protocol performs best when the input phase is in the proximity of the optimal phase given by $\phi_\txt{opt}=0.132\pm0.001$. This is due to the fact that as the input phase diverges from $\phi_\txt{opt}^\txt{th}$, the rough estimation error $\delta\phi_\txt{R}$ gets larger, leading to an increasing error on the phase of feedback corrected probe in comparison to the optimal phase, $|\phi_\txt{opt}^\txt{exp}-\phi_\txt{opt}^\txt{th}|$, which in turn increases the final estimation variance. This is further signified by comparing a non-adaptive with the adaptive approach in Fig.~\ref{Fig_Results} (top).

We further investigate the variance scaling with the number of samples $\text{N}_\txt{tot}$. The experiment is realized for four different values of $\text{N}_\txt{tot}$, and the results are shown in Fig.~\ref{Fig_Results} (bottom) where every data point is an average over several different initial phases distributed across the phase range. We also insert the theoretical predictions of the QCR bounds for the coherent state (orange line) and the pure squeezed state (purple line). From the figure, it is clear that all measurement points are substantially below the SQL (or heterodyne limit) and even below the QCR bound for coherent states.

In conclusion, we are the first to implement a real-time adaptive protocol for ab-initio phase estimation designed to asymptotically saturate the quantum Cramer-Rao limit beyond the shot noise limit. In constrast to a previous implementation of ab-initio phase estimation~\cite{Xiang2010}, we use a source of deterministically generated squeezed states of light and real-time feedback control. We believe that our protocol can be applied to a variety of different metrological and informational tasks, in particular considering the recent advances in the production of pure and highly squeezed states of light~\cite{Eberle2013}.

We acknowledge the financial support of the Danish Research Council (Sapere Aude grant from FTP) and the Lundbeck foundation. The squeezing source was built at the Albert-Einstein institute in Hannover in the group of Prof. R. Schnabel. We would like to thank him for his support.

\newpage

\end{document}